\documentclass[aps, prb, twocolumn, superscriptaddress, letterpaper]{revtex4-2}

\usepackage{graphicx}
\usepackage{amssymb}
\usepackage{amsmath}
\usepackage{txfonts}
\usepackage{bm}
\usepackage{color}
\usepackage{float}
\usepackage[colorlinks=true, linkcolor=blue, anchorcolor=black, citecolor=blue, filecolor=black, menucolor=black, runcolor=black, urlcolor=blue]{hyperref}
\usepackage{orcidlink}

\begin{document}

\title{Three-dimensional flat Landau levels in an inhomogeneous acoustic crystal}

\author{Zheyu Cheng\orcidlink{0000-0001-5009-7929}}
\thanks{These authors contributed equally to this work}
\affiliation{Division of Physics and Applied Physics, School of Physical and Mathematical Sciences, Nanyang Technological University,
Singapore 637371, Singapore}

\author{Yi-jun Guan}
\thanks{These authors contributed equally to this work}
\affiliation{Research Center of Fluid Machinery Engineering and Technology, School of Physics and Electronic Engineering, Jiangsu University, Zhenjiang 212013, China}

\author{Haoran Xue\orcidlink{0000-0002-1040-1137}}
\email{haoranxue@cuhk.edu.hk}
\affiliation{Department of Physics, The Chinese University of Hong Kong, Shatin, Hong Kong SAR, China}

\author{Yong Ge}
\affiliation{Research Center of Fluid Machinery Engineering and Technology, School of Physics and Electronic Engineering, Jiangsu University, Zhenjiang 212013, China}

\author{Ding Jia}
\affiliation{Research Center of Fluid Machinery Engineering and Technology, School of Physics and Electronic Engineering, Jiangsu University, Zhenjiang 212013, China}

\author{Yang Long\orcidlink{0000-0001-7600-3396}}
\affiliation{Division of Physics and Applied Physics, School of Physical and Mathematical Sciences, Nanyang Technological University,
Singapore 637371, Singapore}

\author{Shou-qi Yuan\orcidlink{0000-0003-3252-6608}}
\affiliation{Research Center of Fluid Machinery Engineering and Technology, School of Physics and Electronic Engineering, Jiangsu University, Zhenjiang 212013, China}

\author{Hong-xiang Sun\orcidlink{0000-0003-4646-6837}}
\email{jsdxshx@ujs.edu.cn}
\affiliation{Research Center of Fluid Machinery Engineering and Technology, School of Physics and Electronic Engineering, Jiangsu University, Zhenjiang 212013, China}
\affiliation{State Key Laboratory of Acoustics, Institute of Acoustics, Chinese Academy of Sciences, Beijing 100190, China}

\author{Yidong Chong\orcidlink{0000-0002-8649-7884}}
\email{yidong@ntu.edu.sg}
\affiliation{Division of Physics and Applied Physics, School of Physical and Mathematical Sciences, Nanyang Technological University,
Singapore 637371, Singapore}
\affiliation{Centre for Disruptive Photonic Technologies, Nanyang Technological University, Singapore, 637371, Singapore}

\author{Baile Zhang\orcidlink{0000-0003-1673-5901}}
\email{blzhang@ntu.edu.sg}
\affiliation{Division of Physics and Applied Physics, School of Physical and Mathematical Sciences, Nanyang Technological University,
Singapore 637371, Singapore}
\affiliation{Centre for Disruptive Photonic Technologies, Nanyang Technological University, Singapore, 637371, Singapore}

\begin{abstract}
When electrons moving in two-dimensions (2D) are subjected to a strong uniform magnetic field, they form flat bands called Landau levels, which are the basis for the quantum Hall effect.  Landau levels can also arise from pseudomagnetic fields (PMFs) induced by lattice distortions; for example, mechanically straining graphene causes its Dirac quasiparticles to form a characteristic set of unequally-spaced Landau levels, including a zeroth Landau level.  In three-dimensional (3D) systems, there has thus far been no experimental demonstration of Landau levels or any other type of flat band.  For instance, applying a uniform magnetic field to materials hosting Weyl quasiparticles, the 3D generalizations of Dirac quasiparticles, yields bands that are non-flat in the direction of the field.  Here, we report on the experimental realization of a flat 3D Landau level in an acoustic crystal.  Starting from a lattice whose bandstructure exhibits a nodal ring, we design an inhomogeneous distortion corresponding to a specific pseudomagnetic vector potential (PVP) that causes the nodal ring states to break up into Landau levels, with a zeroth Landau level that is flat along all three directions.  These findings point to the possibility of using nodal ring materials to generate 3D flat bands, to access strong interactions and other interesting physical regimes in 3D.
\end{abstract}

\maketitle

\bigskip
\noindent{\large{\bf{Introduction}}}


Landau levels (LLs) first arose in Landau's 1930 derivation of the magnetic susceptibility of metals, based on a quantum mechanical model of nonrelativistic electrons in a uniform magnetic field \cite{landau1930diamagnetismus}.  If the electrons are constrained to the 2D plane perpendicular to the magnetic field, the LLs form a set of equally-spaced flat energy bands, independent of the in-plane momentum.  Such 2D LLs were subsequently found to exhibit quantized Hall conductance (the integer quantum Hall effect) due to their nontrivial band topology \cite{klitzing1980new, thouless1982quantized}.  Other 2D models host different types of LLs; for example, particles governed by a 2D Dirac equation (such as electrons near the Dirac points of graphene), when subjected to a uniform magnetic field, produce 2D LLs that are flat but unequally spaced in energy, with a zeroth Landau level (0LL) at zero energy \cite{novoselov2005two, zhang2005experimental}. Flat bands such as 2D LLs are of broad interest in multiple fields of physics since their high density of states is conducive for accessing strong-interaction regimes \cite{tsui1982two, laughlin1983anomalous, baba2008slow, yang2023, leykam2018artificial,  rhim2021singular}, such as strong inter-electron interactions in condensed matter systems, which given rise to phenomena like the fractional quantum Hall effect \cite{tsui1982two, laughlin1983anomalous, cao2018unconventional, liu2019pseudo}, and strong light-matter coupling in optoelectronic systems\cite{de2021light, elias2021flat}.  Although LLs are not the only way to achieve flat bands, they are attractive because of their rich physics and relative accessibility.  Aside from using real magnetic fields, LLs can also arise from PMFs induced by lattice distortions, without breaking time-reversal invariance \cite{guinea2010energy}.  This is achievable in electronic materials through strain engineering  \cite{guinea2010energy, levy2010strain, pikulin2016chiral, grushin2016inhomogeneous} or inter-layer twisting~\cite{liu2019pseudo}, and in synthetic metamaterials like photonic or acoustic crystals through structural engineering~\cite{rechtsman2013strain, abbaszadeh2017sonic, yang2017strain, wen2019acoustic, jia2019observation, peri2019axial, bellec2020, jamadi2020direct, wang2020moire, zheng2021landau, yan2021pseudomagnetic, phong2022boundary, cai2023nodal}.  PMFs are highly tunable and can reach much higher effective field strengths than real magnetic fields.

\begin{figure}[b]
  \centering
  \includegraphics[width=\columnwidth]{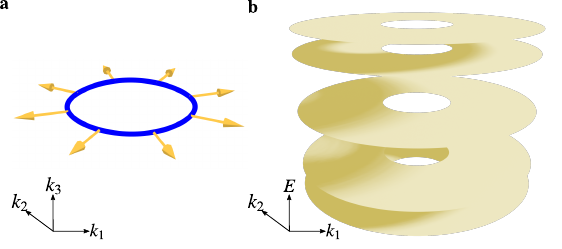}
  \caption{\textbf{Pseudomagnetic field induced Landau levels in 3D nodal ring systems.} \textbf{a}, Illustration of the pseudomagnetic vector potential (yellow arrows) at different positions of the nodal ring (blue circle).  \textbf{b}, Under the pseudomagnetic vector potential  shown in \textbf{a}, the nodal ring splits into 3D flat Landau levels. }
  \label{fig1}
\end{figure}

\begin{figure*}
  \centering
  \includegraphics[width=\textwidth]{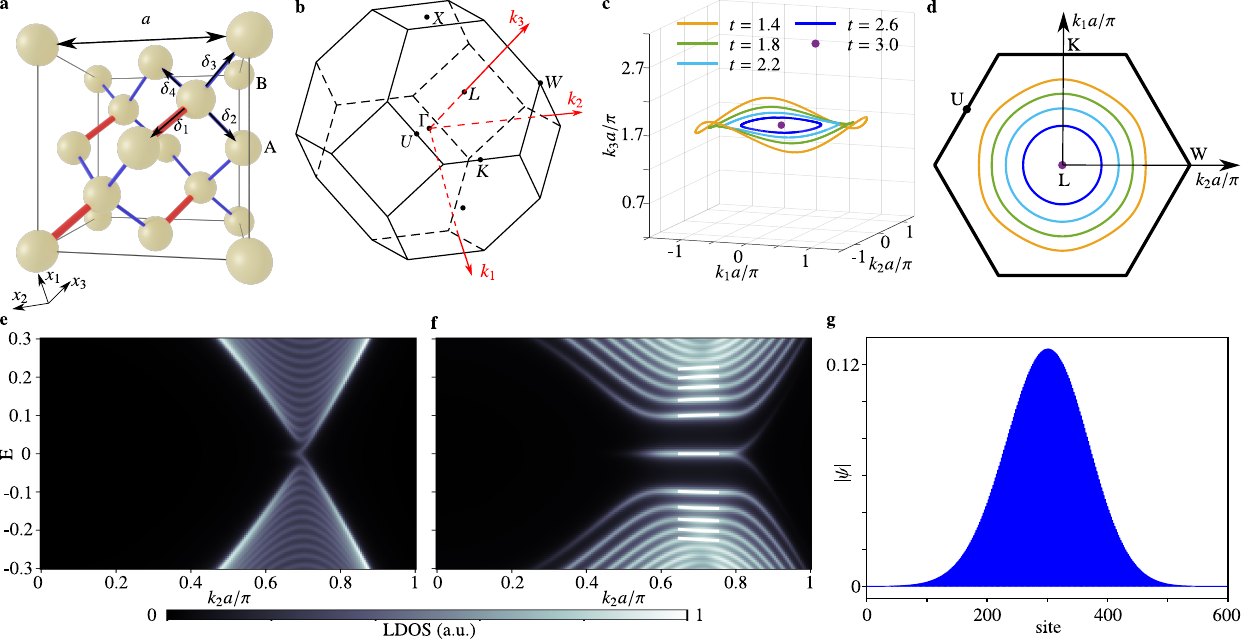}
  \caption{\textbf{3D Landau levels in an inhomogeneous anisotropic diamond lattice.} \textbf{a}, Cubic cell of the anisotropic diamond lattice.  The cubic unit cell has side length $a$.  Red and blue bonds denote nearest-neighbor couplings $t$ and $t' = 1$, respectively, and nearest-neighbor sites are separated by the vectors $\bm\delta_i (i=1,2,3,4)$. The Cartesian coordinate axes are $x_i~(i=1,2,3)$, such that $x_3$ is parallel to $-\bm\delta_1$, and $x_2$ is parallel to $-\bm\delta_3+\bm\delta_4$. \textbf{b}, Schematic of the first Brillouin zone.  The reciprocal lattice vectors  $k_i~(i=1,2,3)$ are oriented along $LK$, $LW$ and $\Gamma L$ respectively. \textbf{c}, Shapes of the nodal ring for various $t$. \textbf{d}, Projections of the nodal ring onto the $k_1$-$k_2$ plane, for the values of $t$ used in \textbf{c}. \textbf{e}-\textbf{f}, Local density of states in the $k_2$ direction for $B = 0$ (\textbf{e}) and $B = 0.0073{a^{ - 2}}$ (\textbf{f}), calculated using a $600$-site chain. Solid white lines in \textbf{f} represent the analytically predicted Landau levels. \textbf{g}, Wavefunction amplitude of the zeroth Landau level at $\left(k_1, k_2 \right) = \left(0, 0.70\pi/a\right)$.}  
  \label{fig2}
\end{figure*}

\begin{figure*}
  \centering
  \includegraphics[width=\textwidth]{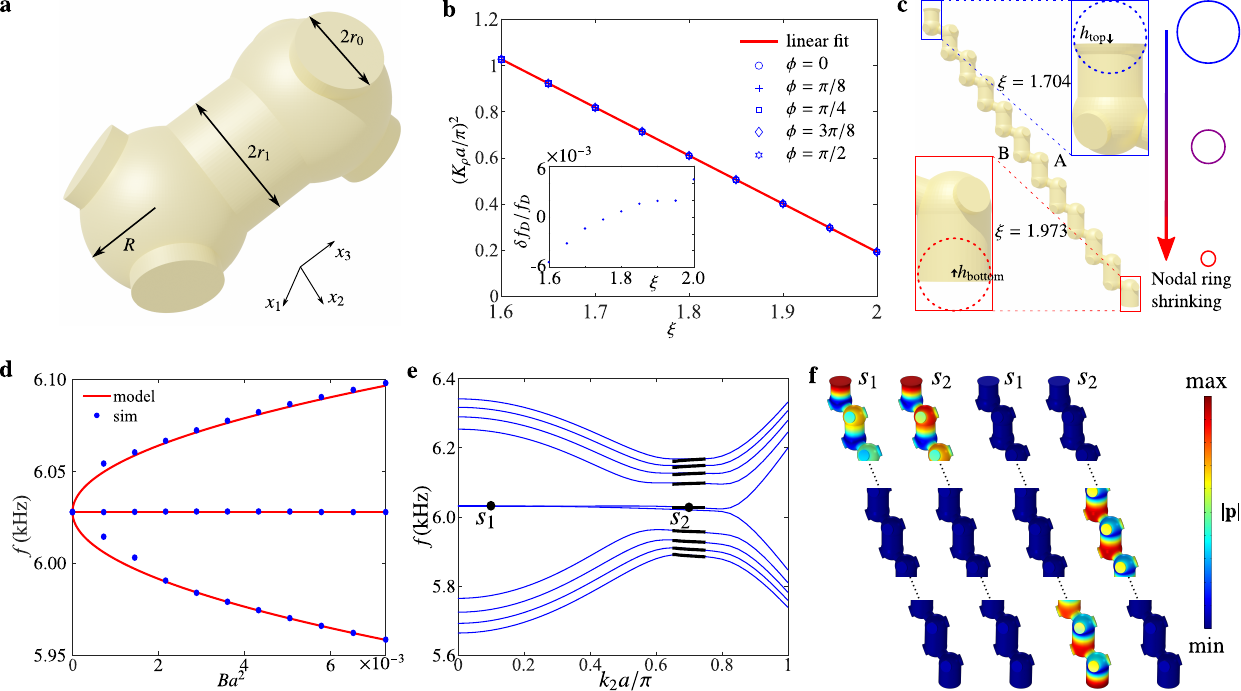}
  \caption{\textbf{Pseudomagnetic fields and Landau levels in an acoustic nodal-ring crystal.} \textbf{a}, Unit cell of the acoustic lattice, consisting of two sphere cavities connected by cylindrical tubes. The radii of the tubes are  $r_1 = \xi r_0$ and $r_0$, respectively. The radius of the sphere cavity is $R$.  \textbf{b}, Plot of the square of the nodal ring's radius $K_\rho = \sqrt{K_1^2+K_2^2}$ against the geometry parameter $\xi$ for different polar angles $\phi$. Blue markers and the red line represent the data and the linear fit, respectively. The lower inset displays the nodal ring's frequency variation when $\xi$ changes. \textbf{c}, Schematic of a 12-layer inhomogeneous acoustic lattice. The value of $\xi$ gradually changes along the $x_3$ direction, which leads to the ring shrinking and induces a pseudomagnetic field. To realize flat drumhead surface states, the top and bottom resonators are cut by $h_\text{top}$ and $h_\text{bottom}$, respectively. \textbf{d}, Eigenfrequencies of the first three Landau levels at $\left( k_1, k_2\right) = \left(0, 0.70\pi/a \right)$ for acoustic lattices under different pseudomagnetic fields. The simulation results (blue dots) are well predicted by the theoretical model (red curves). \textbf{e}, Dispersion along $k_2$ for an inhomogeneous acoustic lattice with 300 layers and $B = 0.0073a^{-2}$. Black lines represent analytically predicted Landau levels. \textbf{f}, Pressure amplitude distributions for the four eigenmodes labelled in \textbf{e}. Both $s_1$ and $s_2$ are double degenerated. One eigenmode localizes at the top surface, whereas another one moves from bottom to bulk as $k_2$ increases. The plots only display the chain's top, middle, and bottom parts and omit other regions where sound pressure is neglectable.}
  \label{fig3}
\end{figure*}

\begin{figure*}
  \centering
  \includegraphics[width=\textwidth]{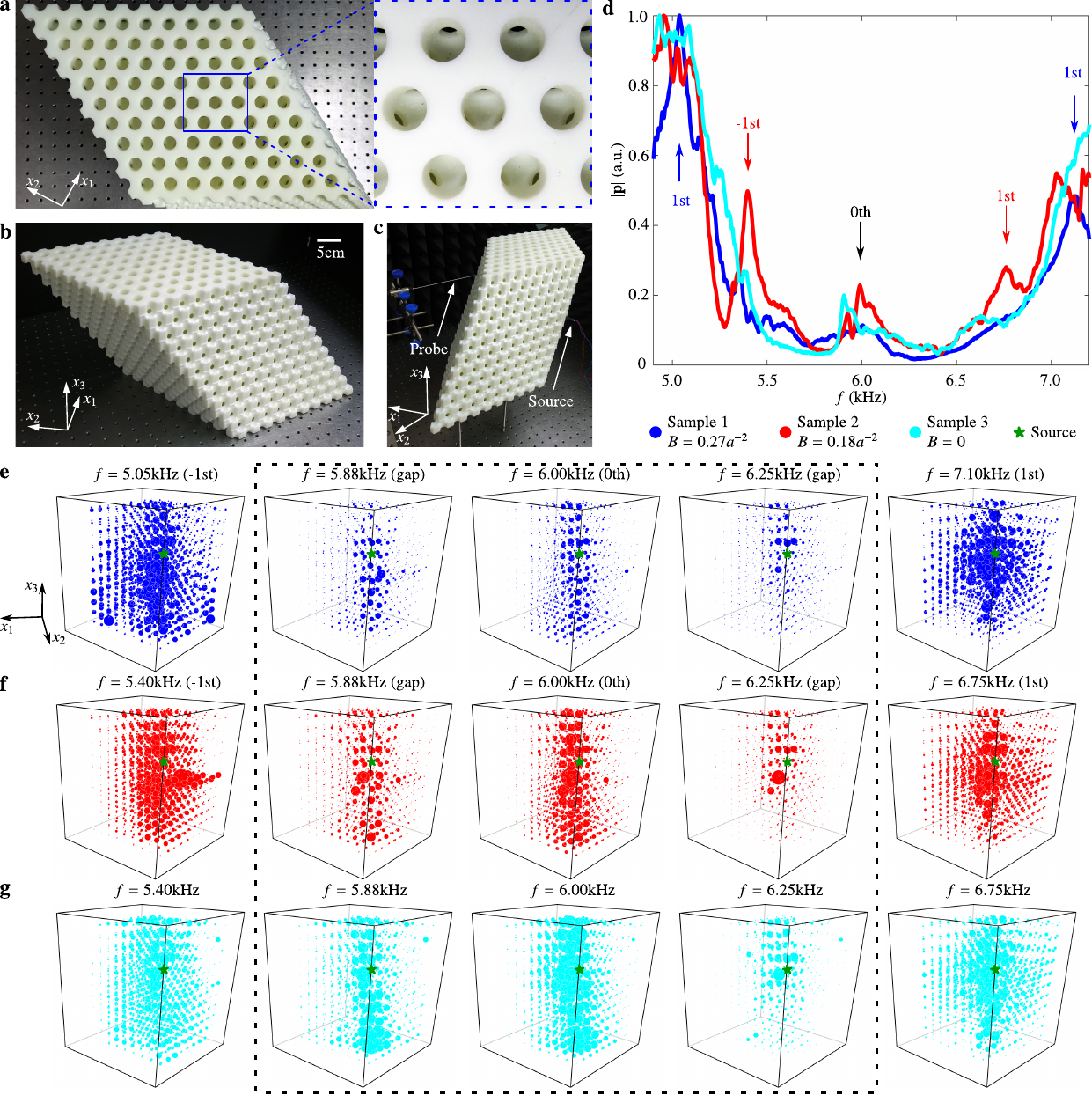}
  \caption{\textbf{Experimental detection of the acoustic Landau levels.}   \textbf{a}, The top view of the sample, with $11\times 11$ sites on $x_1x_2$ plane and 12 layers along $x_3$ direction, which induce strong pseudomagnetic fields under $B = 0.27a^{-2}$, $B = 0.18a^{-2} $, $B = 0$ for sample 1,2,3, respectively.  \textbf{b}, A photo of the sample. Sphere cavities are connected by tubes.  \textbf{c}, The photo of the experiment setup. The probe is ensembled on the mechanical arm, which is controlled by the stepping motor.  \textbf{d}, Measured acoustic pressure spectra at the same bulk site for samples 1 (blue), 2 (red), and 3(cyan). The black arrow indicates frequency $n=0$ Landau levels for sample 1 and 2. The blue (red) arrows indicate frequencies of $n = \{-1, 1 \}$ Landau levels for samples 1(2).  \textbf{e}, Measured acoustic pressure distributions for the $n = \{ -1, 0, 1\}$ Landau levels and two gap frequencies in sample 1. The radii of the blue spheres are proportional to the acoustic pressure.   \textbf{f}, Measured acoustic pressure distributions for the $n = \{ -1, 0, 1\}$ Landau levels and two gap frequencies in sample 2. The radii of the red spheres are proportional to the acoustic pressure.  \textbf{g}, Measured acoustic pressure distributions with five frequencies in sample 3. The radii of the cyan spheres are proportional to the acoustic pressure. The green star denotes the position of the sound source in \textbf{e,f,g}.}
  \label{fig4}
\end{figure*}

In 3D, flat bands are challenging to realize, whether via LLs or some other mechanism; to our knowledge, there is no prior experimental demonstration in any truly 3D structure.  For example, stacking 2D quantum Hall systems turns the LLs into 3D bands that are non-flat along the stacking direction, so long as there is nonzero coupling between layers (similar to the original Landau model)~\cite{tang2019three}.  Likewise, if we generalize 2D Dirac particles to Weyl particles in 3D, applying a uniform magnetic field produces chiral LLs that can propagate freely along the field direction \cite{huang2015observation, armitage2018weyl, yan2023antichirality}.

In this work, we design and experimentally implement a 3D lattice exhibiting LLs that are flat in all three directions.  This is accomplished with an acoustic crystal---a synthetic metamaterial through which classical sound waves propagate---that incorporates an inhomogeneous structural distortion.  In the absence of the distortion, the band structure contains a circular nodal ring (i.e., a ring in momentum space along which two bands touch) \cite{fang2016topological, bzduvsek2016nodal}.  The introduction of the distortion generates a PVP pointing radially from the nodal ring's center in momentum space, as well as varying in real space (Fig.~\ref{fig1}a).  The nodal ring spectrum hence splits into 3D flat bands (Fig.~\ref{fig1}b), in a manner analogous to the formation of 2D LLs from Dirac points \cite{guinea2010energy}.  This 3D flat band spectrum has potential uses for enhancing nonlinearities and accessing interesting 3D phenomena such as inverse Anderson localization \cite{goda2006inverse}.  The acoustic crystal design might also help guide the development of solid-state materials hosting 3D LLs, which could access novel correlated electron phases not found in lower-dimensional flat band systems \cite{lau2021designing}.

\bigskip
\noindent{\large{\bf{Results}}}\\
\noindent\textbf{Continuum model}


In PMF engineering, as performed in strained graphene \cite{guinea2010energy} and related materials \cite{pikulin2016chiral, grushin2016inhomogeneous, peri2019axial, jia2019observation}, a lattice distortion shifts a bandstructure's nodal points (e.g., Dirac points) in momentum space, which is analogous to applying a vector potential $\mathbf{A}$.  For instance, a slowly-varying (compared to the lattice constant) distortion can implement $\mathbf{A} = B x_3 \hat{\mathbf{e}}_1$ (where $x_{1,2,3}$ denotes position coordinates and $\hat{\mathbf{e}}_{1,2,3}$ the unit vectors), corresponding to a uniform PMF $\nabla\times\mathbf{A} = B\, \hat{\mathbf{e}}_2$.  A similar manipulation can be applied to nodal lines, rather than nodal points \cite{rachel2016strain, kim2019elastic, lau2021designing}.  Consider the continuum Hamiltonian \cite{yang2018quantum}
\begin{equation}
  H\left(\mathbf{k}\right)
  = \frac{1}{2m_\rho} \left(k_\rho^2 - k_0^2\right) \sigma _1 + v_3 k_3 \sigma _2,
  \label{continuum_model}
\end{equation}
where $\mathbf{k} = (k_1,k_2,k_3)$ is 3D momentum vector, $\sigma_{1,2}$ are Pauli matrices, $k_\rho^2 = k_1^2 + k_2^2$, and $m_\rho, k_0, v$ are positive real parameters.  This hosts a nodal ring at $k_\rho = k_0$, $k_3 = 0$, with energy $E = 0$ \cite{yang2018quantum}.  Now suppose we impose a PVP
\begin{equation}
  {\bf{A}}\left( {{k_1},{k_2},{x_3}} \right) = B_0 x_3 \hat{\mathbf{e}}_\rho, \label{pseudo-vector_potential}
\end{equation}
where $\hat{\mathbf{e}}_\rho = k_\rho^{-1}(k_1,k_2,0)$ is the radial vector in the plane of the nodal ring.  If we treat $x_3$ as a constant, the Peierls substitution $\mathbf{k} \to \mathbf{k} + \mathbf{A}$ shifts the nodal ring's radius to $k_0' = k_0 -B_0 x_3$.  For a slow variation, $x_3 = - i \partial/\partial k_3$, we can expand $H$ close to the nodal ring (i.e., $\left| k_\rho - k_0' \right| \ll k_0'$) to obtain
\begin{equation}
  H\left(\mathbf{k}\right) \approx \frac{k_0}{m_\rho}
  \left(k_\rho + B_0x_3 - k_0 \right) \sigma _1 + v_3k_3\sigma _2.\label{H_k_A}
\end{equation}
This is a 2D Dirac equation, based on coordinates $\rho$ and $x_3$, with a uniform PMF.  Its spectrum is $E_n = \textrm{sgn}(n) \omega _c\, \sqrt{| n|}$, where $\omega _c = \sqrt{2v_3 B_0 k_0 / m_\rho}$ and $n = 0, 1, 2, \dots$ (see Supplementary Information).  Each LL is flat along $k_\rho$, $k_3$, as well as the nodal ring plane's azimuthal coordinate (which $H$ does not depend upon).

\noindent\textbf{Lattice model}

Following our scheme, the key to realizing 3D LLs is to have a bandstructure with a circular nodal ring, whose radius can be parametrically varied without losing its circularity.  Such a situation arises in a tight-binding model of an anisotropic diamond lattice \cite{takahashi2013completely}.  As shown in Fig.~\ref{fig2}a, the cubic unit cell has period $a$, there are two sublattices with one $s$ orbital per site, and the nearest-neighbor couplings are $t$ (red bonds) and $t'$ (blue bonds).  The momentum-space lattice Hamiltonian is
\begin{equation}
H(\bf{k})= \left( {\begin{array}{*{20}{c}}
  0&{te^{i\bf{k}\cdot\bm\delta_1} + t'\sum\nolimits_{i = 2}^4 {{e^{i{\bf{k}}\cdot{{\bm{\delta}}_i}}}} } \\ 
  {{te^{-i\bf{k}\cdot\bm\delta_1}} + t'\sum\nolimits_{i = 2}^4 {{e^{ - i{\bf{k}}\cdot{{\bm{\delta}}_i}}}} }&0 
\end{array}} \right),
\label{eq1}
\end{equation}
where $\bm\delta_1, \dots, \bm\delta_4$ are the nearest-neighbor displacements shown in Fig.~\ref{fig2}a (see Supplementary Information).  The first Brillouin zone is depicted in Fig.~\ref{fig2}b. Thereafter, we set $t'=1$ for convenience. This lattice is known to host a nodal ring when $1<t<3$ \cite{takahashi2013completely}, whereas for $t>3$ it is a higher-order topological insulator \cite{ezawa2018minimal, xue2019realization}. In the former regime, the nodal ring occurs at $E = 0$  and
\begin{align}
K_1^2 + K_2^2 &=\left(\frac{2\sqrt{2}}{a}\sqrt{3-t}\right)^2, \label{loop-shape} \\
K_3&= \sqrt{3}\frac{\pi}{a},
\end{align}
where $\left(K_1, K_2, K_3 \right)$ is the position on the nodal ring.  The nodal ring forms a circle in momentum space (see Supplementary Information Fig.~S1, and Figs.~\ref{fig2}c--\ref{fig2}d). Crucially, its radius is determined solely by $t$.  We can form any shape of nodal lines in the continuum model \cite{burkov2011topological}, but it is hard to realize circle-shaped nodal rings. In most cases nodal line semimetals hold either discrete lines \cite {bzduvsek2016nodal} or closed rings with other shapes \cite{kim2015dirac, rhim2015landau, bzduvsek2016nodal, xiong2018topological, deng2019nodal}.

To generate the 3D LLs, we modulate $t$ along the spatial axis $x_3$ perpendicular to the plane of the ring, so that the nodal ring's radius increases linearly with $x_3$.  This leads to the gauge field (see Supplementary Information):
\begin{equation}
{{\bf{A}}\left( x_3, \phi\right)} = B x_3 \left( {\begin{array}{*{20}{c}}
  {\cos \phi } \\ 
  {\sin \phi } \\
  {0}
\end{array}} \right), 
\label{eq3}
\end{equation}
with parameter $B = {0.4}{\sqrt 3 \pi}/{ Na^2}$, which controls the magnitude of PVP. Here, $N$ is the number of unit cells along the $x_3$ direction, and $\phi$ is the azimuth angle in the $k_1$--$k_2$ plane.  Such a PVP splits the original nodal degeneracy into discrete LLs described by ${E_n} = \operatorname{sgn} \left( n \right)\sqrt {\left| n \right|} {\omega _c}$, where ${\omega _c} = {v}\sqrt {2 B } $ is the cyclotron frequency (see Figs.~\ref{fig2}e--\ref{fig2}f). While the nonzero LLs are not ideally flat due to the $\bf{k}$-dependence of the group velocity $v$, the zeroth LL is exactly dispersionless, which indicates a novel mechanism for generating flat bands in 3D systems. A numerically calculated profile of the zeroth LL is plotted in Fig.~\ref{fig2}g, whose localization in the bulk is well-captured by the low-energy theory (see Supplementary Information).

\begin{figure}[t]
  \centering
  \includegraphics[width=\columnwidth]{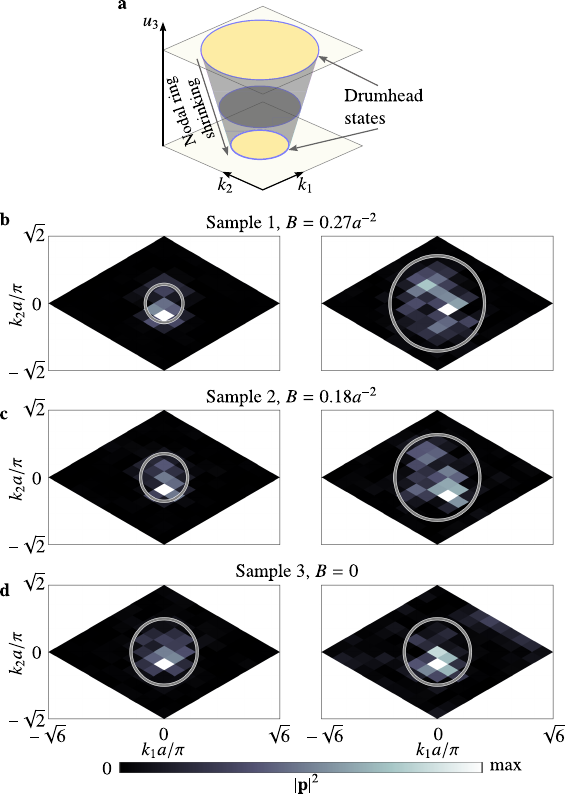}
  \caption{\textbf{PMF-modified drumhead surface states.} \textbf{a}, Illustration of the drumhead states at the top and bottom surface. \textbf{b,c,d} Measured Fourier transformed intensity at the bottom (top) surface at 6.12kHz for three samples. The gray circles denote the projections of the nodal ring near the surfaces.}
  \label{fig5}
\end{figure}

\noindent\textbf{Acoustic lattice design}

Next, we design an acoustic diamond lattice to realize the above phenomenon. Figure~\ref{fig3}a shows the acoustic unit cell, which consists of two sphere cavities connected by cylindrical tubes with radii $r_1$ or $r_0$. The whole structure is filled with air and covered by hard walls. Here, the two sphere cavities act as two sublattices (denoted as ``A" and ``B"; see Fig.~\ref{fig3}c) and the cylindrical tubes control the coupling strengths. We define a dimensionless parameter $\xi = r_1/r_0$ \, which describes the anisotropic strength of couplings. Similar to the tight-binding model, this acoustic lattice hosts a circular-like nodal ring in its band structure. We numerically find that the square of the radius of the nodal ring is controlled by a single parameter $\xi$ and it scales linearly with $\xi$ (see Fig.~\ref{fig3}b). Moreover, for different values of $\xi$, the frequency of the nodal ring remains almost unchanged (see the inset to Fig.~\ref{fig3}b). We can straightforwardly engineer PMFs in this acoustic lattice with these nice properties. The supercell of the designed structure is schematically illustrated in Fig.~\ref{fig3}c, where $\xi$ gradually varies along the $x_3$ direction. This variation makes the radius of the nodal ring linearly dependent on space coordinates, similar to the case in the tight-binding model. We cut part of the top and bottom cavities to tune on-site potential. After precisely controlling $h_\text{top}$ and $h_\text{bottom}$, the drumhead surface state is almost dispersionless, which is the advantage in the drumhead state measurement. 

We numerically test our design using a lattice with 300 layers along the $u_3$ direction. Figure~\ref{fig3}d plots the frequencies of $n=\{-1,0,1\}$ LLs under different PMFs. As can be seen, the spacings between the LLs follow well with the theoretical curve ${E_n} = \operatorname{sgn} \left( n \right)\sqrt {\left| n \right|} {\omega _c}$. Figure~\ref{fig3}e shows the dispersion along $k_2$ direction under $ B= 0.0073a^{-2}$, where the acoustic LLs is clearly presented. Notably, the zeroth LL is exponentially localized in the bulk and is smoothly connected to the surface modes (see Fig.~\ref{fig3}f). All these numerical results are consistent with the low-energy theory and tight-binding calculations.

\noindent\textbf{Experiment}

To observe the LLs experimentally, we fabricate three 12-layer samples with stronger PMFs under $B = 0.27a^{-2}$, $B = 0.18a^{-2} $, $B = 0$, as shown in Figs.~\ref{fig4}a--\ref{fig4}c. Figure~\ref{fig4}a shows the top layer of the sample. Figure~\ref{fig4}c shows the experiment setup. The mechanical arm is controlled by the stepping motor, which ensures we get the pressure field in the right position. The strong PMF leads to LLs with spacing that is wide enough to be measured in a simple pump-and-probe experiment (see Methods for more experimental details). Figure~\ref{fig4}d plots the measured spectrum when the source and the probe are placed at two bulk sites (see Supplementary Information Fig.~S8). As can be seen, there are pronounced peaks at the corresponding frequencies of the predicted LLs. 

To visualize the effect of the PMF, we plot the acoustic field distributions at several representative frequencies that correspond to the frequencies of the $n=\{-1,0,1\}$ LLs or the middle frequencies between these LLs. As shown in Fig.~\ref{fig4}e--\ref{fig4}f, the excited fields spread a noticeable area when the source operates at LL frequencies. In contrast, the excited fields are highly confined to the source position at midgap frequencies. 

Furthermore, we compare the acoustic field distributions at Dirac frequency and gap frequencies for three samples. As shown in the 2nd to the 4th columns in Fig.~\ref{fig4}e-\ref{fig4}g, the stronger the PMF, the more localized the field. Such a sharp comparison is a direct consequence of the Landau quantization of the acoustic bands. 

Conventional nodal line crystals are accompanied by the drumhead surface states. In the presence of the PMF, we find that the drumhead surface states are also modified. As illustrated in Fig.~\ref{fig5}a, due to the spatial variation of the nodal ring's radius, the momentum space area of the drumhead surface states at the top and bottom surfaces are different (see Supplementary Information). To see this effect, we measure the acoustic fields at the top and bottom surfaces, and the corresponding Fourier spectra are given in Fig.~\ref{fig5}b--\ref{fig5}d. Due to the size effect, the drumhead surface state frequency is located at 6.12kHz, shifting a little from the zeroth LL. As shown in Figs.~\ref{fig5}b--\ref{fig5}c, the surface states at the top surface indeed occupy a larger area in the momentum space compared to those at the bottom.

\bigskip
\noindent{\large{\bf{Discussion}}}

To sum up, we have theoretically proposed and experimentally demonstrated the generation of PMFs in 3D nodal-ring systems. Our results open a new route to studying the physics of artificial gauge fields in gapless systems beyond Dirac and Weyl semimetals. From the perspective of wave manipulation, the PMF-induced LLs provide a novel method to generate flat bands in 3D systems \cite{leykam2018artificial}, which could be useful in sound trapping, energy harvest, and slow-wave devices. In future studies, it would be interesting to investigate the effects of other forms of PMF beyond the constant one \cite{phong2022boundary} and the interactions between PMFs and other types of band degeneracies, such as nodal link, nodal knot, and nodal surfaces. Extending the idea to photonic and electronic systems is also highly desired, where nonlinear and correlated physics can be studied.

\bigskip
\noindent{\large{\bf{Methods}}}

\noindent\textbf{Tight-binding calculations.} The nodal ring structures in Figs.~\ref{fig2}c--\ref{fig2}d are obtained by directly diagonalizing the lattice Bloch Hamiltonian (i.e., Eq.~\eqref{eq1}) and then tracing the degenerate eigenvalues. The local density of states of plots (Figs.~\ref{fig2}e--\ref{fig2}f) are calculated via Green's function:
\begin{equation}
G = \frac{1}{E+i\gamma -H}
\end{equation}
as $ - \text{Im} \left( {\sum\nolimits_{i \in \text{400 bulk sites}} {{G_{ii}}} } \right)/\pi $.  Here $H$ is the Hamiltonian for a supercell that has 600 sites along $x_3$ direction and is periodic in the other two directions and $\gamma = 0.01$. In Fig.~\ref{fig2}e, we fix $t$ unchanged and the radius of the nodal ring is $0.70\pi/{a}$. In Fig.~\ref{fig2}f, The coupling parameter $t$ varies along $x_3$, and the radius of the nodal ring expands from $0.5\pi/a$ (bottom) to $0.9\pi/a$ (top).

\noindent\textbf{Acoustic lattice design.} The side length of the cubic cell is set to be $a=4$ cm, while other structural parameters are all tunable. All tunable structural parameters are controlled by the dimensionless parameter $\xi$. Specifically, the radii of the tubes are given by $r_1 = \xi r_0$ and $r_0$ with $r_0=0.4$ cm, and the radius of the sphere is $R=0.8$ cm. 

\noindent\textbf{Numerical simulations.} All simulations are performed using the acoustic module of COMSOL Multiphysics, which is based on the finite element method. The photosensitive resin used for sample fabrication is set as the hard boundary due to its large impedance mismatch with air. The real sound speed at room temperature is $c_0 = 346\text{m/s}$. The density of air is set to be 1.8 kg/m$^3$. The results in Fig.~\ref{fig3}b are obtained by computing the bandstructure of one unit cell, with the Floquet boundary condition applied to all directions. The data points are selected by scanning the dispersion along $k_1, k_2, k_3$ and tracing the degenerate points. We note this is a good measure of the nodal ring's radius due to its circular shape (see Fig.~\ref{fig3}b and Supplementary Information Fig.~S5). In the simulations of Figs.~\ref{fig3}d--\ref{fig3}f, a supercell with 300 layers along the $x_3$ direction is used. In Figs.~\ref{fig3}e--\ref{fig3}f, $\xi$ changes from $1.973$ to $1.704$ from bottom to top. Accordingly, the radius of the nodal ring expands from $0.5\pi/a$ to $0.9\pi/a$. To realize dispersionless drumhead surface states, the top and bottom resonators are cut by $h_\text{top} = 0.246 r_0$ and $h_\text{bottom} = 0.216 r_0$, respectively. More simulations can be found in Supplemental Information.

\noindent\textbf{Sample details.} The sample is fabricated via 3D printing technology with a  fabrication error of 0.1 mm. The triclinic sample has a side length is 28.28 cm $\times$ 28.28 cm $\times$ 32.01 cm and contains 2904 sphere cavities and several waveguides.  All spheres at the surface are cut in half to mimic radiation boundary conditions. During the measurement, the sample is surrounded by sound-absorbing sponges to reduce the reflection from the sample boundaries. When we measure the top (bottom) drumhead surface states, the top (bottom) surface is covered by an acrylic plate (Supplementary Information Fig.~S10). All three sample has 12 layers.  The parameters of three samples are listed in table.~\ref{sample}.

\begin{table}[H]
\centering
\caption{Parameters of three samples.}
\label{sample}
\begin{tabular}{|c|c|c|c|c|}
\hline
Sample	&	$\xi$	&	$k_\rho a /\pi$	&	$h_\text{top}$	&	$h_\text{bottom}$	\\
\hline
1	&	1.612 to 2.016	&	1.0 to 0.4	&	$0.258 r_0$	&	$0.148 r_0$	\\
\hline
2	&	1.704 to 1.973	&	0.9 to 0.5	&	$0.246 r_0$	&	$0.216 r_0$	\\
\hline
3	&	1.858	&	0.7	&	$0.278 r_0$	&	$0.226 r_0$	\\
\hline
\end{tabular}
\end{table}

\noindent\textbf{Experimental measurements.} In the LL experiment, a broadband sound signal (4 kHz to 8 kHz) is launched from a narrow long tube in Fig.~\ref{fig4}c (diameter of about 0.35 cm and length of 35 cm) that is inserted into the centermost site, which acts as a point-like sound source for the wavelength focused here. The pressure of each site is detected by a microphone (Brüel\&Kjær Type 4961) adhered to a long tube (of diameter about 0.35 cm and a length of 35 cm). The signal is recorded and frequency-resolved by a multi-analyzer system (Brüel\&Kjær 3160-A-022 module). In the drumhead state experiment, the source is located at the 2nd (12th) layer when we measure the bottom (top) drumhead state (see Supplememtary Information Fig.~S10). Other sets are the same as before.

\noindent\textbf{Data analysis.} In the LL experiment, pressure data of the outermost sites (cut sites and sites that connect them directly) are discarded. All pressure data are normalized by the source's frequency spectrum before they are used to create the plots in Fig.~\ref{fig4}. In the drumhead state experiment, all data are processed by Fourier transformation.

\bigskip
\noindent{\large{\bf{Data availability}}}

\noindent The experimental data are available in the data repository for Nanyang Technological University at this link (\href{https://doi.org/10.21979/N9/RS60NB}{https://doi.org/10.21979/N9/RS60NB}). Other data supporting this study's findings are available from the corresponding authors upon reasonable request.

\bigskip
\noindent{\large{\bf{Acknowledgements}}}

Z.C., Y.L., and B.Z. are supported by the Singapore Ministry of Education Academic Research Fund Tier 2 Grant MOE2019-T2-2-085, and National Research Foundation (NRF), Singapore under its Competitive Research Programmes NRF-CRP23-2019-0007. Y.C. is also supported by the National Research Foundation (NRF), Singapore under its Competitive Research Programmes NRF-CRP23-2019-0005, and NRF Investigatorship NRF-NRFI08-2022-0001. Y.G., Y.G., D.J., S.Y., and H.S. acknowledge support from the National Natural Science Foundation of China under Grant Nos.12274183 and 12174159, and the National Key Research and Development Program of China under Grant No. 2020YFC1512403. H.X. acknowledges support from the start-up fund of The Chinese University of Hong Kong. We thank Daniel Leykam for helpful comments.

\bigskip
\noindent{\large{\bf{Author contributions}}}

\noindent H.X. conceived the idea. Z.C., H.X., and Y.L. did the theoretical analysis. Z.C. performed the simulations and designed the sample. Y-J. G., Y.G., D.J., and H.S. conducted the experiments. Z.C., H.X., S.Y., Y.C., and B.Z. wrote the manuscript with input from all authors. H.X., H.S., Y.C., and B.Z. supervised the project.

\bigskip
\noindent{\large{\bf{Competing interests}}}

\noindent The authors declare no competing interests.
\end{document}